# Magnetic Confinement by Alfven Carpets

A. B. Hassam
*University of Maryland, College Park, MD 20742*

*Abstract*
A "carpet" of torsional Alfven waves, resonantly excited in an annulus inside a toroidal conducting vessel, could maintain magnetized plasma inside of the annulus against the inherent outward expansion. Such toroidal magnetic confinement for fusion could present some advantages over conventional tokamaks. The idea could also be used to stabilize interchange and ballooning modes in tokamaks and other interchange-limited fusion devices.

Magnetic confinement for fusion starts from the idea that a 10 keV proton in a 1 T magnetic field is constrained to gyrate the field at a radius of 1 cm. The difficulty is in containing the plasma in the third dimension, along the field. The naive idea would be to make the field toroidal, creating a bicycle-tire plasma flux tube. As is well known though[1,2], the tire immediately expands radially outward at the sound speed, roughly 1000 km/s, as there is no topological frozen-in MHD constraint to the outward expansion (Fig 1).

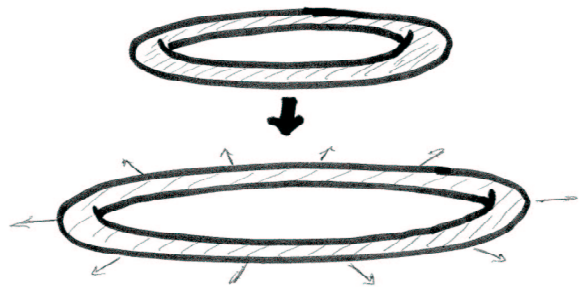
Fig. 1

The tokamak solution is to add a poloidal twist to the purely toroidal field, introducing a topological constraint for the nested surfaces toward outward expansion (Fig 2). The combination of the required toroidal and poloidal fields, however, necessitates linked toroidal and poloidal field coils, making for the engineering complexity of the system. Alternative means of providing a frozen-in constraint to the outward expansion are thus worth exploring.

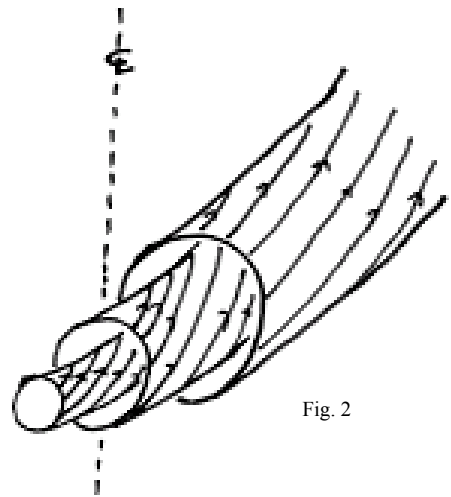
Fig. 2

One such possibility is discussed in this paper. Consider a conducting toroidal vacuum vessel with a toroidal magnetic field, $B_T$. Let there be plasma in the vessel. We now resonantly excite torsional Alfven waves on the outer rim of the vessel (Fig. 3). The plasma inside of the Alfven wave ``carpet'' on the outer rim will be prone to the outward expansion. The shear Alfven carpet, however, will prevent any outward expansion beyond the carpet radius, on account of the frozen-in condition. An estimate shows that the required poloidal field amplitude of the Alfven wave, $B_P$, must be

$$(B_P/B_T)^2 \;>\; \varepsilon\beta/6, \qquad\qquad (1)$$

where $\beta = 3nT/[B^2/2\mu_0]$ and $\varepsilon = a/R$ is the inverse aspect ratio.

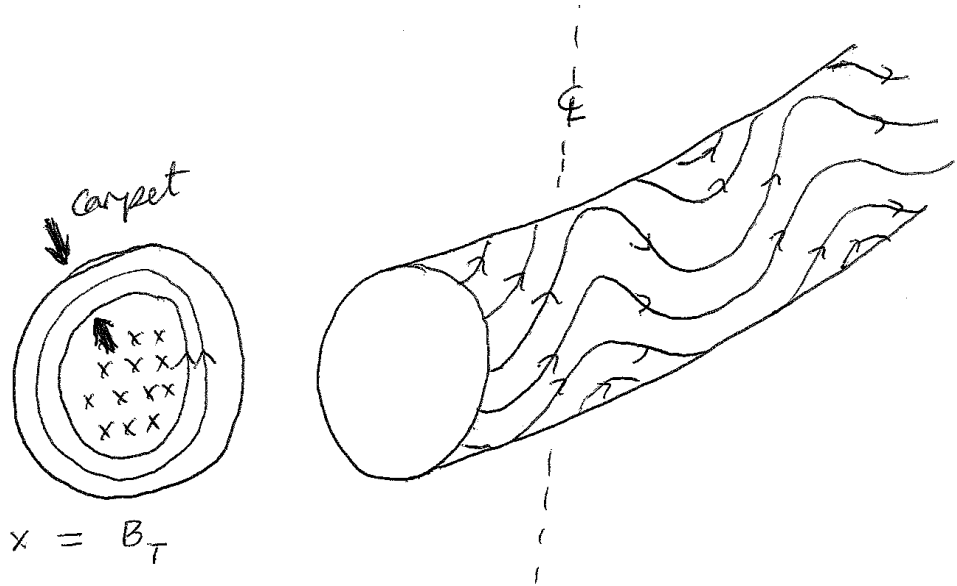

The system is also likely to be stable to interchanges beyond the carpet radius. Based on a recent calculation[3], we estimate that

$$(B_P/B_T)^2 > \varepsilon\beta/24 \qquad (2)$$

might suffice.

At resonance, the required wave power is $P \sim 2(B_P^2/2\mu_0)(Vol)(2/f)/\tau_\mu$, where $1/\tau_\mu = (\mu/a^2)$ is the ion crossfield viscous damping rate of the plasma and f is the fraction of the radius a that makes up the carpet thickness. We are assuming that the Alfven wave envelope will dissipate viscously. We compare the required wave power with that needed to reach energy breakeven, $P_L = (3nT/\tau_E)(Vol)$ with T = 10 keV and $\tau_E$ being the energy confinement time, ie, we estimate the circulating power fraction needed to maintain the carpet. We find

$$P/P_L \sim (2\varepsilon/3f)\, \tau_E/\tau_\mu \qquad (3)$$

where we have used (1). This fraction is independent of $\beta$. It is likely that the viscous time and the energy confinement time may be comparable though the former may be somewhat longer. Thus, this amounts to significant circulating power, upto $\sim 2/3$ fraction. The viscous losses will heat the plasma, though, reducing auxiliary power requirements.

The main concern is Ohmic losses in the vacuum vessel; such losses were found to be a limiting issue for a related confinement system investigated earlier[4]. The resonant frequency for traveling waves is $\omega = V_A/R$. Equating the wave period of (1/30 Mhz) with $\tau_R$, we calculate the penetration depth $\Delta$ of the waves into a copper shell to be

$$\Delta(m) \sim 2 \cdot 10^{-4} [R(m)/B(T)]^{1/2} n_{20}^{1/4}$$

for a copper shell with resistivity $\eta_{Cu} = 17$ n$\Omega$ m. The resistive penetration time in copper over a scale size $\Delta$ is given by $\tau_{R,Cu} = 60 \Delta(m)^2$ sec. The wall loading due to the Ohmic losses in the wall can now be estimated as[4]

$$P_{wall}/A \sim (B_P^2/2\mu_0) \Delta/\tau_{R,Cu} \sim 5 \epsilon\beta B_T(T)^{5/2} R(m)^{-1/2} n_{20}^{-1/4} \quad MW/m^2.$$

For $B_T = 1$ T, $\beta = 1$, and $\epsilon = 1/3$, this amounts to $\sim 2$ MW/m$^2$.

Finally, in a similar vein, Alfven carpets could be set up in present day tokamaks, to enhance ballooning $\beta$ limits[1], or in other interchange-limited fusion concepts[2]. A calculation of interchange growth mitigation from Alfven carpets is described in Ref 3. Tokamak experiments could be an intermediate step toward a testing of this concept. Stabilization of interchange modes by ponderomotive forces from radio-frequency fields in magnetic mirrors and in tokamaks has previously been studied in Ref 5.

The Alfven carpet confinement system can be run steady state, and at $\beta \sim 1$ for strong enough wave amplitudes. There might be no disruptions. The circulating power fraction may be significant. Alpha particles will be confined. We have not yet investigated the practicality of suitable antenna systems to resonantly drive the Alfven waves. These might present difficulty. If feasible, the antenna and the toroidal field coil system may have lesser overall linkages compared to tokamak poloidal and toroidal field coil systems.

We acknowledge useful discussion with Prof R. J. Goldston.